\documentclass[12pt]{article} 			%\usepackage{bezier,epic,cite}
\usepackage{amssymb}  %\topmargin=-15mm

\usepackage{amsmath}
\usepackage{latexsym,cite}		%	\catcode`\"=\active \let"=\"
\def\({\left(}       \def\){\right)}	\let\6=\partial \let\S=\Sigma
\let\3=\ss
\newcommand{\vs}[1]{\vspace{#1 mm}}

\renewcommand{\thefootnote}{\fnsymbol{footnote}}
\begin{document}
\begin{titlepage}
\setcounter{page}{0}
\begin{flushright}
hep-th/0005148\\
TUW-00-15 
\end{flushright}				\def\new#1\endnew{{\bf #1}}

\vs{10}
\begin{center}
	{\Large\bf $SU(2)$ WZW D-branes and quantized worldvolume $U(1)$	flux on $S^2$}
\vs{15}

{\large   Alexander KLING\footnote{e-mail: kling@@hep.itp.tuwien.ac.at}, 
Maximilian KREUZER\footnote{e-mail: kreuzer@hep.itp.tuwien.ac.at}
and Jian-Ge ZHOU\footnote{e-mail: jgzhou@hep.itp.tuwien.ac.at}}\\
\vs{10}
{\em  Institut f\"ur Theoretische Physik, Technische Universit\"at Wien,
        Wiedner Hauptstra\3e  8--10, A-1040 Wien, AUSTRIA}\\
\end{center}

\vs{6}

\vs{15}
\centerline{{\bf{Abstract}}}
\vs{5}

We discuss possible D-brane configurations on $SU(2)$ group manifolds in the 
sigma model approach. When we turn the boundary conditions of the spacetime 
fields into the boundary gluing conditions of chiral currents, we find that 
for all D-branes except the spherical D2-branes, the gluing matrices 
$R^a_{~b}$ depend on the fields, so the chiral Kac-Moody symmetry is broken,
but conformal symmetry is maintained. 
Matching the spherical D2-branes derived from the sigma model with those from 
the boundary state approach we obtain a $U(1)$ field strength that is
consistent with flux quantization.

% can be determined 
% and the resulting $U(1)$ worldvolume flux has to be quantized.

\end{titlepage}
\newpage
\renewcommand{\thefootnote}{\arabic{footnote}}
\setcounter{footnote}{0}

\section{Introduction}

In recent years there has been much interest in the study of D-branes on 
group manifolds (see for instance \cite{1,2,3,4,5,6,7}). String theory on 
group manifolds is governed by a WZW model, which 
has two distinct descriptions: % conformal field theory (CFT) 
the current algebra and the sigma model realization. 
Since the WZW model is a typical example of an exact string background, whose 
CFT is known explicitely, one approach to find possible D-brane configurations
is to impose gluing conditions on the chiral currents 
$J^a(z)$ and $\bar J^a(\bar z)$ in terms of which the CFT is defined. 
Actually the boundary state approach has been applied widely to find D-brane 
configurations on group manifolds \cite{2,3,4,6,7,A,B}. In \cite{4} 
it was found that the D-branes in WZW models associated with the gluing 
condition $J^a = -\bar J^a$ along the boundary are configurations of 
quantized conjugacy classes. As the gluing conditions in the 
boundary state approach are defined on the chiral currents rather than on 
the spacetime fields there is a lack of an obvious geometric interpretation 
of WZW boundary states and in particular of the corresponding 
D-brane configurations. Since the WZW model provides also an example of a 
string background with a sigma model description, which allows a 
complementary study of the D-brane configurations, it is interesting 
to compare the D-brane configurations obtained from the sigma model 
realization with those from the boundary state approach (CFT) in order to 
see how they match with each other. 

The other motivation for this work was to see the quantization of the 
worldvolume $U(1)$ flux on the spherical D2-brane. 
In \cite{8,9} it was suggested that the $U(1)$ worldvolume flux $\int F$ 
rather than that of $\int [(2\pi\alpha')^{-1} B + F]$ should be quantized. 
In \cite{8,9}, the quantization problem was mainly discussed from the
Born-Infeld theory, so it is quite interesting to see whether we can study it
from the worldsheet perspective. Since the $U(1)$ gauge field 
appears in the action of the sigma model, we wonder what the D-brane 
configurations constructed from the sigma model approach
has to say about this problem.

Motivated by the above, in this paper we study WZW D-branes on the group 
manifold of $SU(2)$ from the sigma model point of view
% We construct D0-, D1-, D2-, and D3-branes by the sigma model approach, then 
and compare the results % ing D-branes 
with % those from 
the boundary state approach. Our strategy is that we turn 
the boundary conditions of the spacetime fields into gluing condition of the 
chiral current at the boundary %, for  D2-, D3-branes we 
and try to adjust the $U(1)$ gauge field to make the gluing matrices 
field independent in order to check chiral Kac-Moody symmetry. 
For the spherical D2-branes we find that in order to keep the 
infinite-dimensional symmetry of the current algebra, 
the $U(1)$ worldvolume gauge field strength has to take the form 
$F=-\frac{\kappa}{2\pi} \psi_{0} \epsilon_2$,
where $\kappa$ is the integer level of the associated current algebra and 
$\psi_{0}$ describes the radius of the spherical D2-branes. 
Imposing the quantization condition on the flux $\int F$ that follows from
a consistent definition of the sigma model action we will recover the 
quantization of the brane positions $\psi^{(n)}_{0} = n\pi/\kappa$ \cite{4,5}.
%From the general Bohr-Sommerfeld quantization condition \cite{4,5}, the radius
%of the spherical D2-branes is quantized, i.e. with the values 
%$\psi^{(n)}_{0} = n\pi/\kappa$. By matching the 
%spherical D2-branes in the CFT approach to those in the complementary sigma 
%model, the $U(1)$ worldvolume flux $\int F$ has to be quantized. 
For other D-branes we find it impossible to adjust the $U(1)$ gauge field to 
make the gluing matrices $R^a_{~b}$ field independent (the gluing matrix is 
defined by the gluing condition $J^a(z) + R^a_{~b} \bar J^a(\bar z) =0$ at 
the boundary). The dependence of the gluing 
matrices $R^a_{~b}$ on the spacetime fields certainly breaks the chiral 
Kac-Moody symmetry, but we find that conformal invariance is maintained 
for these D-branes, at least in our classical approximation.
The straightforward extension our result to twisted conjugacy classes 
is briefly discussed in the last section.

\section{Parametrization of the sigma model action and chiral currents} 

We start with the $SU(2)$ WZW action with gauge % fields
bundles $A$ defined on the brane subspaces of the group manifold
% into which the boundaries of the world sheets $\S$ are to be embedded
\cite{1}:
\begin{equation}\label{sigma}
	S = \frac{\kappa}{8\pi}\int_{\Sigma} % d^2\sigma \,
	tr(g^{-1}\partial_+ g g^{-1}\partial_- g)
	 + \frac1{2\pi\alpha'}\int_{\Sigma}g^* B + \int_{\partial\Sigma} g^* A,
\end{equation}
where we use the string normalization of target space fields.
For world sheets with boundary the WZ part of the action is defined in 
terms of the field strengths $H=dB$ and $F=dA$ by employing a closed 
auxiliary surface $\S'=\6M$ that is the union of $\S$ with $n$ disks $D^{(i)}$
if the boundary $\6\S$ has $n$ connected components, and by extending $g$ to 
the 3-dimensional manifold $M$:
\begin{equation}	S_{WZ} = \frac{\kappa}{12\pi}\int_{M} 
	tr(g^{-1}d g)^{3} - \sum_{i=1}^n\frac{1}{2\pi\alpha'}\int_{D^{(i)}}
	g^* (B + 2 \pi\alpha'F).
\end{equation}
The $U(1)$ field strength $F=dA$ is defined on the D-brane submanifold
(i.e. at the allowed positions of the boundaries of the embedded
world sheet; in case of several branes we have to introduce an independent
gauge connection $A$ for every brane and the respective field strength has to
be used for each disk). Moreover, we need to choose a gauge where $B$ is not
singular on the brane.

Independence of all quantum amplitudes of the various 
choices involved in this definition implies that the level $\kappa$ 
and all integrals $\int_{S^2} F/2\pi$ of the respective field strengths 
$F$ over any 2-spheres ${S^2}$ embedded in the branes have to 
be integers \cite{4,C}: If we close some component of the boundary of $\S$ with
two different disks $D$ and $D'$, then the difference in the action is
$\frac1{2\pi\alpha'}((\int_{M'}-\int_{M})H -(\int_{D'}-\int_{D})
	(B+2\pi\alpha' F))$. Since $H=dB$ globally on the respective brane,
	$B$ drops out and we are left with an integral of $F$ over 
	the 2-sphere $D\cup D'$.

We can also think about the action (\ref{sigma}) in the following way:
The gauge transformation $B\to B+d\Lambda$ leads to surface terms that 
can only be compensated if we introduce a gauge field $A$ that transforms
as $A\to A-2\pi\alpha' \,\Lambda$ at the boundary. The gauge symmetry 
$A\to A+d\lambda$ just corresponds to the trivial part of the reducible 
$\Lambda$-transformation that leaves $B$ invariant. Hence only the field 
strength $H$ is physical outside the branes, whereas at the allowed positions 
of the boundary of the world sheet $F+B/2\pi\alpha'$ also becomes observable.

%where $\Sigma$ is the $2D$ manifold with boundary $\partial\Sigma$, $B$ is a 
%particular choice for the 
%antisymmetric tensor field, and the overall factor in (\ref{sigma}) has been
%omitted. Here we note that
%for the $SU(2)$ group manifold, the B-field in (\ref{sigma}) is defined only 
%locally
% \footnote{The proper global form for 
% Wess-Zumino term is $S = \frac{\kappa}{12\pi}\int_{\Sigma + D^2} 
% tr(g^{-1}\partial g)^{3} - \frac{\kappa}{12\pi}\int_{D^2}g^* (B + 2 \pi 
% \alpha'F)$, where $\Sigma + D^2$ has no boundary, and the disc $D^2$
% is mapped to D-brane submanifold \cite{5}.}.
 
% In the case of $SU(2)$ 
To proceed we choose the parametrization 
\begin{equation}\label{paramet}
g= \begin{pmatrix} \cos \psi - i \sin \psi \sin \theta \sin \phi & 
                   - \sin\psi\sin\theta\cos\phi-i\sin\psi\cos\theta \\
                   \sin\psi\sin\theta\cos\phi-i\sin\psi\cos\theta & 
                   \cos \psi + i \sin \psi \sin \theta \sin \phi \end{pmatrix}
\end{equation}
of the group manifold with
\begin{equation}
   0 \le \psi \le \pi, \quad  0 \le \theta \le \pi, \quad 0 \le \phi \le 2\pi.
\end{equation}
In these coordinates the metric and the NS three-form field are given by
\begin{eqnarray}
  & ds^2 = \kappa\alpha'[d\psi^2+\sin^2\psi(d\theta^2+\sin^2\theta d\phi^2)]
% \end{equation}\begin{equation}
  &\\[5pt]	
&	H =  \frac{1}{6} \kappa \alpha' \textrm{tr}(g^{-1}dg)^3 
	= 2 \kappa \alpha' \sin^2 \psi \sin\theta d\psi \wedge d\theta 
	\wedge d\phi	&		\label{dh}
\end{eqnarray}
% where $\kappa$ is the integer level of the associated current algebra. 
% Then the $SU(2)$ WZW 
and the action turns into 
\begin{eqnarray}\label{action}
  S & = & \frac1{2 \pi \alpha'}
	 \int d\tau d\sigma \{\frac{1}{2} \eta^{\alpha\beta} \kappa \alpha' 
  \(\partial_{\alpha}\psi\partial_{\beta}\psi + \sin^2 \psi 
  \partial_{\alpha}\theta\partial_{\beta}\theta + 
  \sin^2 \psi \sin^2 \theta \partial_{\alpha}\phi\partial_{\beta}\phi\) 
	\nonumber
\\
  & &+ B_{\theta\phi}(\partial_{\tau}\theta\partial_{\sigma}\phi - 
  \partial_{\sigma}\theta\partial_{\tau}\phi) \nonumber
%\\& &
  + B_{\psi\theta}(\partial_{\tau}\psi\partial_{\sigma}\theta - 
  \partial_{\sigma}\psi\partial_{\tau}\theta) \nonumber
\\
  & &+ B_{\psi\phi}(\partial_{\tau}\psi\partial_{\sigma}\phi - 
  \partial_{\sigma}\psi\partial_{\tau}\phi) \}+ 
  \int_{\partial\Sigma} g^* A
\end{eqnarray}
where $\eta^{\alpha\beta}=\textrm{diag}(-1,1)$. 

The WZW model has conserved chiral currents 
(with $\partial_{\pm} = \partial_{\tau} \pm \partial_{\sigma}$),
\begin{equation}\label{currents}
J = - \partial_{+}g g^{-1}, \qquad \bar J = g^{-1}\partial_{-}g 
\end{equation}
Inserting the parametrization (\ref{paramet}) into (\ref{currents}) we have 
\begin{equation}\label{current1}
   J^a = - \bar e^a_{~\mu}\partial_{+}X^{\mu},\qquad \bar J^a =  e^a_{~\mu}
	\partial_{-}X^{\mu}
\end{equation}
\begin{equation}\label{vielbein1}
\bar e^a_{~\mu} = -\begin{pmatrix} 
\cos\theta & & - \sin\psi\cos\psi\sin\theta & & \sin^2\psi\sin^2\theta 
\\
& &  & &
\\
\sin\theta\cos\phi & & \sin\psi\cos\psi\cos\theta\cos\phi & & 
-\sin\psi\cos\psi\sin\theta\sin\phi 
\\
 & & -\sin^2\psi\sin\phi & & - \sin^2\psi\sin\theta\cos\theta\cos\phi 
\\
 & &  & &
\\
\sin\theta\sin\phi & & \sin\psi\cos\psi\cos\theta\sin\phi & & 
\sin\psi\cos\psi\sin\theta\cos\phi 
\\
 & & + \sin^2\psi\cos\phi & & - \sin^2\psi\sin\theta\cos\theta\sin\phi
\end{pmatrix} 
\end{equation}
\begin{equation}\label{vielbein2}
e^a_{~\mu} = \begin{pmatrix} 
-\cos\theta & & \sin\psi\cos\psi\sin\theta & & \sin^2\psi\sin^2\theta 
\\ 
& &  & &
\\
- \sin\theta\cos\phi & & - \sin\psi\cos\psi\cos\theta\cos\phi & & 
\sin\psi\cos\psi\sin\theta\sin\phi 
\\
& & - \sin^2\psi\sin\phi & & - \sin^2\psi\sin\theta\cos\theta\cos\phi
\\
& &  & &
\\
- \sin\theta\sin\phi & & - \sin\psi\cos\psi\cos\theta\sin\phi & & 
- \sin\psi\cos\psi\sin\theta\cos\phi 
\\
 & &  + \sin^2\psi\cos\phi & &  - \sin^2\psi\sin\theta\cos\theta\sin\phi
\end{pmatrix} 
\end{equation}
where $X^1=\psi,X^2=\theta,X^3=\phi$, and $J^a,\bar J^a$ are defined by 
\begin{equation}
J^a = \frac{i}{2} \textrm{tr}(\sigma^a\partial_{+}g g^{-1}), \qquad 
      \bar J^a = - \frac{i}{2} \textrm{tr}(\sigma^a g^{-1}\partial_{-}g)
\end{equation}
and $\sigma^a$ are the Pauli matrices. 
The vielbein matrices $e$ and $\bar e$ satisfy 
$e^T e = \bar e^T \bar e = G$, where $G$ is the metric
\begin{equation}\label{metric}
G_{\mu\nu} = \textrm{diag}(1,\sin^2 \psi, \sin^2\psi\sin^2\theta)  .
\end{equation}

\section{D-brane configurations constructed in the sigma model approach}

When we vary the action (\ref{action}) we get the equations of motion.
In addition we work out the 
boundary conditions from which we can construct possible D-brane 
configurations. %	\footnote{The action (\ref{action}) contains
%	$U(1)$ gauge field $A$ only at the boundary $\partial\Sigma$, 
%	however when we vary the action (\ref{action}), the boundary
%	part has the sort of $B+2\pi\alpha' F$ term.}
We first consider the solution
% \footnote{Here we choose $B_{\theta\phi}= \kappa \alpha' 
% 	(\psi - \frac{\sin 2\psi}{2})\sin\theta$, and the other 
% 	choice of B will be considered below.} 
\begin{equation}\label{bf}
	(B+2\pi\alpha' F)_{\theta\phi} 
	= \kappa \alpha' ( \psi - \frac{\sin 2\psi}{2} + f)\sin\theta
\end{equation} 
to $H=dB+2\pi\alpha' F$ with constant $f$ and with all other components 
vanishing, which is suggested by the symmetry of our choice of coordinates.%
\footnote
{	Two other simple choices are
%	\begin{eqnarray}\label{choiceI}	% I) & & 
	$(B + 2\pi\alpha' F)_{\psi\theta} 
	= 2 \kappa\alpha'(\phi\sin^2\psi\sin\theta + f')$ 
%	\\\label{choiceII} % II) & & 
and $(B + 2\pi\alpha' F)_{\psi\phi} 
	= 2 \kappa\alpha'(\sin^2\psi\cos\theta + f'')$ % \end{eqnarray}	
	with all other components vanishing in both cases. 
	For $f\neq0$ they are, however, singular at $\theta=0$ mod $\pi$ 
	for all $\psi$ and therefore not useful for $D2$ branes.
}
The 2-form $\sin\theta d\theta d\phi$ is singular at $\psi=0$ and at 
$\psi=\pi$, which suggests to associate the term proportional to $f$ with the 
gauge field strength $F$. The remaining $B$ field is then regular everywhere 
except at the point $\psi=\pi$.
% zero at the boundary $\partial\Sigma$. Here the form of $B_{\theta\phi}$ is 
% smooth everywhere except at the point $\psi=\pi$, as $\kappa$ is integer 
% this potential singularity is unobservable \cite{8} and 
% $F_{\theta\phi} = \frac{\kappa}{2\pi} f\sin\theta$ in
% D-brane submanifold\footnote{The ends of open string are
% sensitive to the concrete choice of B and F, the bulk
% of string feels only the NS 3-form field H, so $F\neq 0$
% only on the D-brane submanifold.}.
% As $dF=0$ we demand that on the D-brane worldvolume $f$ should be constant. 
% Since we have the freedom of changing $B$ by an exact 2-form, which is 
% nothing but the $U(1)$ gauge field strength $F$, the quantity 

% First we restrict our attention to branes that are contained in conjugacy 
% classes so that $\psi=\psi_0$ is constant. 
At the present stage of our discussion $f$ is an undetermined parameter. 
We can formaly extend the domain of $F$ and effectively obtain a family of 
choices for the $B$ field.
Obviously $f\to f-\pi$ then shifts the $B$ field into another gauge with a
singularity in a single point, but this time at $\psi=0$, the unit element
of the group (at the same time the flux $\int F/2\pi$ through a sphere at 
fixed $\psi=\psi_0$ is shifted by the integer $-\kappa$).
It turns out that $f$ can not be fixed by the leading order condition of 
conformal invariance at the boundary given in~\cite{3} 
\begin{equation}
\partial_{\mu}[\sqrt{G} G^{\mu\nu}G^{\rho\sigma}(B+2\pi\alpha' F)_{\nu\rho}]=0
\end{equation}
where the metric is given by (\ref{metric}).

With the choice % of $(B+2\pi\alpha' F)_{\theta\phi}$ in 
(\ref{bf}) we can read off the boundary 
condition by varying the action (\ref{action}) and we find
\begin{eqnarray}\label{bndcond}
	(\delta\,\psi\partial_{\sigma}\psi)\Big{|}_{\partial\Sigma}  
	& = & 0 \nonumber
\\
	\delta\theta \(\sin^2\psi\,\partial_{\sigma}\theta - 
	(\psi - \frac{\sin 2\psi}{2} + f)\sin\theta\,\partial_{\tau}\phi\) 
	\Big{|}_{\partial\Sigma} & = & 0 \nonumber
\\
	\delta\phi \(\sin^2\psi\sin^2\theta\,\partial_{\sigma}\phi + 
	(\psi - \frac{\sin 2\psi}{2} + f)\sin\theta \partial_{\tau}\theta\) 
	\Big{|}_{\partial\Sigma}  & = & 0 
\end{eqnarray}
By exploiting (\ref{bndcond}) we can look for D-brane 
% construct D0-, D1-, D2- and D3-brane 
configurations of various dimensions by considering the
following simple boundary conditions. \\
D0-brane:
\begin{equation}\label{d0}
\psi\Big{|}_{\partial\Sigma} = \psi_0,\quad \theta\Big{|}_{\partial\Sigma} = 
	\theta_0, \quad \phi\Big{|}_{\partial\Sigma} = \phi_0
\end{equation}
D1-branes:
\begin{eqnarray}\label{d1}
  & & \psi\Big{|}_{\partial\Sigma} 
	= \psi_0,\quad \theta\Big{|}_{\partial\Sigma} = \theta_0,
	\quad \partial_{\sigma}\phi\Big{|}_{\partial\Sigma} = 0
\\
  & & \psi\Big{|}_{\partial\Sigma} 
	= \psi_0,\quad \partial_{\sigma}\theta\Big{|}_{\partial\Sigma} = 0,
	\quad \phi\Big{|}_{\partial\Sigma} = \phi_0
\\ \label{d1'}
  & & \partial_{\sigma}\psi\Big{|}_{\partial\Sigma} 
	= 0,\quad \theta\Big{|}_{\partial\Sigma} = \theta_0,
	\quad \phi\Big{|}_{\partial\Sigma} = \phi_0
\end{eqnarray}
Spherical D2-brane:\footnote{When $\psi_0=0$ and $\pi$, the 
spherical D2-branes reduce to D0-branes, and  the D0-branes
described by (\ref{d0}) can be derived from the D0-branes
of $\psi_0=0$ and $\pi$ by an inner automorphism.}
\begin{eqnarray}\label{sd2}
	\psi\Big{|}_{\partial\Sigma} &=& \psi_0 \nonumber
\\
	\(\sin^2\psi\,\partial_{\sigma}\theta - 
	(\psi - \frac{\sin 2\psi}{2} + f)\sin\theta\,
	\partial_{\tau}\phi\)\Big{|}_{\partial\Sigma} &=& 0 \nonumber
\\
	\(\sin^2\psi\sin^2\theta\,\partial_{\sigma}\phi + 
	(\psi - \frac{\sin 2\psi}{2} + f)\sin\theta \partial_{\tau}\theta\) 
	\Big{|}_{\partial\Sigma}&=& 0
\end{eqnarray}
where $\psi_0,\theta_0,\phi_0$ are arbitrary constants.
The last two D1-brane candidates still have to be closed by the 
antipodal halfs of the respective circles but we have to drop them from
our considerations anyway because our ansatz for $B+F$ is singular at their
location (this is not a big loss, however, because these configurations can 
be obtaind from global rotations of the group manifold, as we will
discuss below).
Replacing the Dirichlet boundary condition in (\ref{sd2}) by 
% a van Neumann one 
$\partial_{\sigma}\psi\big{|}_{\partial\Sigma} = 0$ we formally
get a D3-brane, but this is inconsistent because we cannot have a B-field
without singularity on the group manifold.

\section{Comparison of the D-brane configurations between two approaches and 
	quantized $U(1)$ worldvolume flux on $S^2$}

Now we compare the D-brane configurations derived from the above sigma model 
with those from the boundary 
state approach. To do so, we construct the gluing condition 
$J^a(z) + R^a_{~b} \bar J^a(\bar z)\big{|}_{\partial\Sigma} =0$ from the 
boundary condition of the spacetime 
fields $\psi,\theta,\phi$ for various D-brane configurations. We try to 
adjust the undetermined parameter 
$f$ to see whether we can get spacetime field independent gluing matrices 
$R^a_{~b}$ in order to check 
the infinite-dimensional symmetry of the current algebra. 

For the following comparison, we need the explicit
expressions for $J^a$ and $\bar J^a$. Using (\ref{current1})-(\ref{vielbein2}) 
we rewrite them as 
\begin{eqnarray}\label{gluing1}
   J^1 &=& \cos\theta\partial_{\tau}\psi+\cos\theta\partial_{\sigma}\psi-
   \sin\psi\cos\psi\sin\theta\partial_{\sigma}\theta+
   \sin^2\psi\sin^2\theta\partial_{\tau}\phi \nonumber
\\
  & & +\sin\theta(\sin^2\psi\sin\theta\partial_{\sigma}\phi-
  \sin\psi\cos\psi\partial_{\tau}\theta) \nonumber
\\
  J^2 &=& \sin\theta\cos\phi\partial_{\tau}\psi + 
  \sin\theta\cos\phi\partial_{\sigma}\psi  - 
  \sin^2\psi\sin\phi\partial_{\tau}\theta + 
  \sin\psi\cos\psi\cos\theta\cos\phi\partial_{\sigma}\theta \nonumber
\\
& & -\sin^2\psi\sin\theta\cos\theta\cos\phi\partial_{\tau}\phi -
  \sin\psi\cos\psi\sin\theta\sin\phi\partial_{\sigma}\phi \nonumber
\\
& & - \sin\phi(\sin^2\psi\partial_{\sigma}\theta 
	+ \sin\psi\cos\psi\sin\theta\partial_{\tau}\phi )\nonumber
\\
  & & - \cos\theta\cos\phi(\sin^2\psi\sin\theta\partial_{\sigma}\phi - 
  \sin\psi\cos\psi\partial_{\tau}\theta) \nonumber
\\
  J^3 &=&  \sin\theta\sin\phi\partial_{\tau}\psi + 
  \sin\theta\sin\phi\partial_{\sigma}\psi  + 
  \sin^2\psi\cos\phi\partial_{\tau}\theta + 
  \sin\psi\cos\psi\cos\theta\sin\phi\partial_{\sigma}\theta \nonumber
\\
  & & -\sin^2\psi\sin\theta\cos\theta\sin\phi\partial_{\tau}\phi +
  \sin\psi\cos\psi\sin\theta\cos\phi\partial_{\sigma}\phi \nonumber
\\
  & & + \cos\phi(\sin^2\psi\partial_{\sigma}\theta 
  + \sin\psi\cos\psi\sin\theta\partial_{\tau}\phi )\nonumber
\\
  & & - \cos\theta\sin\phi(\sin^2\psi\sin\theta\partial_{\sigma}\phi - 
  \sin\psi\cos\psi\partial_{\tau}\theta) 
\end{eqnarray}
\begin{eqnarray}\label{gluing2}
  \bar J^1 &=& -\cos\theta\partial_{\tau}\psi+\cos\theta\partial_{\sigma}\psi-
  \sin\psi\cos\psi\sin\theta\partial_{\sigma}\theta+
  \sin^2\psi\sin^2\theta\partial_{\tau}\phi \nonumber
\\
  & & -\sin\theta(\sin^2\psi\sin\theta\partial_{\sigma}\phi-
  \sin\psi\cos\psi\partial_{\tau}\theta) \nonumber
\\
  \bar J^2 &=& - \sin\theta\cos\phi\partial_{\tau}\psi + 
  \sin\theta\cos\phi\partial_{\sigma}\psi - 
  \sin^2\psi\sin\phi\partial_{\tau}\theta + 
  \sin\psi\cos\psi\cos\theta\cos\phi\partial_{\sigma}\theta \nonumber
\\
  & & -\sin^2\psi\sin\theta\cos\theta\cos\phi\partial_{\tau}\phi -
  \sin\psi\cos\psi\sin\theta\sin\phi\partial_{\sigma}\phi \nonumber
\\
  & &  + \sin\phi(\sin^2\psi\partial_{\sigma}\theta 
	+ \sin\psi\cos\psi\sin\theta\partial_{\tau}\phi ) \nonumber
\\
  & &  + \cos\theta\cos\phi(\sin^2\psi\sin\theta\partial_{\sigma}\phi - 
	\sin\psi\cos\psi\partial_{\tau}\theta) \nonumber
\\
  \bar J^3 &=& - \sin\theta\sin\phi\partial_{\tau}\psi + 
  \sin\theta\sin\phi\partial_{\sigma}\psi  + 
  \sin^2\psi\cos\phi\partial_{\tau}\theta + 
  \sin\psi\cos\psi\cos\theta\sin\phi\partial_{\sigma}\theta \nonumber
\\
  & & -\sin^2\psi\sin\theta\cos\theta\sin\phi\partial_{\tau}\phi +
  \sin\psi\cos\psi\sin\theta\cos\phi\partial_{\sigma}\phi \nonumber
\\
  & & - \cos\phi(\sin^2\psi\partial_{\sigma}\theta 
	+ \sin\psi\cos\psi\sin\theta\partial_{\tau}\phi )\nonumber
\\
  & & + \cos\theta\sin\phi(\sin^2\psi\sin\theta\partial_{\sigma}\phi - 
  \sin\psi\cos\psi\partial_{\tau}\theta)
\end{eqnarray}
Now let us first consider the spherical D2-brane characterized by (\ref{sd2}). 
In the boundary state approach the spherical D2-brane is described
by the gluing condition \cite{4}
\begin{equation}\label{glucond}
  J^a = \bar J^a
\end{equation}
at the boundary $\partial\Sigma$. Here we % should notice that we 
have turned the gluing 
condition for the spherical D2-brane in the boundary state
approach, which uses the closed string picture, into the % that in 
open string picture\footnote{There should be a minus sign difference 
between open and closed string picture \cite{6}.}. 
Comparing $J^a$ and  $\bar J^a$ we find that consistency of
% to match the spherical D2-brane described by 
the boundary conditions (\ref{sd2}) % in sigma model to that  described by 
with the gluing codition (\ref{glucond}) in the boundary state approach,
% we have to demand
requires\footnote{For example, let us consider
$J^1 = \bar J^1$, the first line of $J^1$ is equal to that of
$\bar J^1$ with the help of the first equation in (\ref{sd2}),
but the second line differs a minus sign. To get $J^1 = \bar J^1$
at the boundary $\partial\Sigma$, we must demand 
$(\sin^2\psi\sin\theta\partial_{\sigma}\phi - 
\sin\psi\cos\psi\partial_{\tau}\theta)\Big{|}_{\partial\Sigma} = 0$.
When we exploit the third equation in (\ref{sd2}), we obtain (\ref{sd2cond}).}
\begin{equation}\label{sd2cond}
  \psi_{0} - \frac{\sin 2\psi_{0}}{2} + f = - \sin\psi_{0}\cos\psi_{0},
\end{equation}
which results in 
\begin{equation}\label{ff}
  f=-\psi_{0}.
\end{equation}
In \cite{4} it was shown that the D-brane configurations in the WZW model 
associated with the gluing condition 
$J^a = - \bar J^a$ (in the closed string picture) are the conjugacy classes, 
and in the case of $SU(2)$ group 
the D-brane configurations are spherical D2-branes, which are described 
by the boundary conditions (\ref{sd2}) in the sigma model approach.
\def\del#1\enddel{} 
\del 
To see how the position of the spherical D2-brane is 
quantized, let us recall that the action  (\ref{sigma}) can
be derived from the WZW action. Since a closed loop on $S^{2}$ can
be contracted in distinct ways which gives rise the
ambiguity\footnote{Since the Wess-Zumino term is $\frac{\kappa}{12\pi}
\textrm{tr}(g^{-1}dg)^3$, comparing
with (\ref{dh}) we have the factor $1/2\pi\alpha'$. }
\begin{equation}\label{qc}
\Delta I=\frac{1}{2\pi\alpha'}\(\int_{M}H - \int_{S^{2}}
(B+2\pi\alpha'F)\) 
\end{equation}
and we require it to be the value of $2\pi n$ with integer n, where M is 
one of the 3-balls bounded by the conjugacy class $S^{2}$ \cite{4,5}.
Inserting (\ref{dh}), (\ref{bf}) and (\ref{ff})
into (\ref{qc}) we 
have\footnote{In deriving (\ref{result}), we exploit that
$\int_{M}H = \int d\Omega_{2}\int^{\psi_{0}}_{0}2\kappa
\alpha´\sin^{2}\psi d\psi = 4\pi\kappa\alpha´({\psi_{0}}
-\frac{\sin2\psi_{0}}{2}),$ and $\int_{S^2}(B+2\pi\alpha'F)
 = \kappa\alpha´\int d\Omega_{2}(-\frac{\sin2\psi_{0}}{2})
 = -2\pi\kappa\alpha´\sin2\psi_{0}$.}
\begin{equation}\label{result}
\psi^{(n)}_{0} = \frac{n\pi}{\kappa}
\end{equation}
which indicates that the $n$-th sphere locates at $\frac{n\pi}{\kappa}$. 
Then the $U(1)$ worldvolume gauge field strength 
on the n-th spherical D2-brane is $F=-\frac{\kappa}{2\pi}\, 
\psi^{(n)}_{0}\epsilon_2 = - \frac{n}{2} \,\epsilon_2$. So the $U(1)$ 
worldvolume flux $\int F$ is quantized 
$(-2\pi n)$\footnote{When we replace
(\ref{bf}) by $(B+2\pi\alpha' F)_{\theta\phi} = \kappa \alpha' ( \psi -\pi
- \frac{\sin 2\psi}{2} + \tilde{f})\sin\theta$, the singular
point is transfered to $\psi = 0$. By the same procedure we get
$\tilde{f}=\pi - \psi_0$, and the U(1) worldvolume flux $\int F = 2(\kappa-n)
\pi$.}, which supports the
hypothesis in \cite{8}. 
\enddel
Imposing the quantization of the $U(1)$ 
worldvolume flux $\int F/2\pi=f\kappa/\pi$ that follows from the 
definition of the action we thus recover the quantization 
% \begin{equation}\label{result}
$ \psi^{(n)}_{0} = \frac{n\pi}{\kappa}$
% \end{equation}
of the brane positions. %, which verifies the hypothesis in \cite{8}. 

\del
What we have learned from the above is that if we require the spherical 
D2-brane configuration derived from the 
sigma model to match that from the boundary state approach, the $U(1)$ 
worldvolume gauge field strength has 
to be fixed\footnote{Certainly, there is $\Lambda$-transformation 
under which $F$ is not invariant, but it does not affect $\int F$~\cite{8}.}. 
Here we emphasize that because 
of $dF=0$ the parameter $f$ should be constant on the 
D-brane worldvolume. For the spherical D2-brane we have $f=-\psi_{0}$ which 
is consistent with its boundary 
condition $\psi \big{|}_{\partial\Sigma} = \psi_0$.
\enddel

For D0- and D1-brane configurations the gluing condition can be written as
\begin{equation}
J^a + R^a_{~b} \bar J^b =0
\end{equation}
with 
\begin{equation}
 R = \bar e y e^{-1},
\end{equation}
where the vielbein matrices $e$ and $\bar e$ are defined in (\ref{vielbein1},
\ref{vielbein2}) and the matrix $y$ is 
defined by
\begin{equation}
\partial_+ X^\mu = -y ^\mu_{~\nu} \partial_- X^\nu.
\end{equation}
For the D0-branes $y=\textrm{diag}(1,1,1)$, so that $R$ corresponds to the
inner automorphism that translates the brane to the unit at $\psi=0$ \cite 6.
For D1-branes at constant $\psi$ and $\theta$, on the other hand, we find 
$y=\textrm{diag}(1,1,-1)$.
% are $\textrm{diag}(1,1,-1),\textrm{diag}(1,-1,1),\textrm{diag}(-1,1,1)$ 
% respectively.

Since there is no place to put the magnetic field strength 
that could balance the tension on a D1-brane worldvolume, 
the D1-brane configurations are believed to be unstable. 
\del For instance, the D1-brane cycle with 
the boundary condition (\ref{d1}) will shrink to a point like object which 
forms a nonmarginal bound state with the stable spherical D2-brane. 
\enddel 
Except for the case of spherical D2-branes, and trivially for the D0 branes,  
% for all other D-branes 
the gluing matrices $R^a_{~b}$ depend on the %spacetime fields
target space position, which indicates that the chiral Kac-Moody symmetry 
is broken. For the $SU(2)$ group manifold, the energy-momentum tensor is 
$T(z)=\frac{1}{\kappa + 2} J^a J^a$. 
Since $R^T R = 1$, we have $T(z)=\bar T(\bar z)$ at the 
boundary, so that conformal invariance is preserved even 
though the chiral Kac-Moody symmetry is broken~\cite{3}.
% which shows that in the presence of the D0-,D1-branes even 
% though the chiral Kac-Moody symmetry 
% is broken, the theory still preserves conformal invariance. 

\section{Summary and discussion}

% In the above, 
We have investgated possible D-brane configurations from the 
sigma model point of view. In order to see what 
the counterparts of these D-branes are in the boundary state approach, we 
turned the boundary conditions of the 
spacetime fields into  gluing conditions of the chiral currents at the 
boundary. We have shown that except 
for spherical D2-brane configurations the gluing matrices for all other 
D-brane configurations depend on the 
spacetime fields. For the spherical D2-branes we have seen that 
the configurations derived from the sigma model do not match those from the 
boundary state approach 
automatically. If we demand that they coincide with each other, the 
$U(1)$ worldvolume flux $\int F$ has to be quantized as has been conjectured
in \cite{8}, and as it indeed follows from the ambiguity in the definition of
the action.
\del
we have to 
put a strong restriction on the form of the 
$U(1)$ gauge field strength. Actually, the gauge field strength can be 
determined by such a matching, which indicates 
that open strings are quite sensitive to not only the concrete
choice of the 2-form potential B but also the $U(1)$ gauge
field strength F on the D-brane worldvolume . Furthermore  we have found 
that the $U(1)$ worldvolume flux $\int F$ has to be quantized, which 
supports the hypothesis in \cite{8}. 
\enddel

\del
In (\ref{bf}) we considered the special choice for the field 
$B + 2\pi\alpha' F$ at the boundary $\partial\Sigma$ from which we constructed 
spherical D2-branes and others. One may ask how about the other two choices 
for the NS B-field, where the nonzero  
$B + 2\pi\alpha' F$ takes the following forms respectively
\begin{eqnarray}\label{choiceI}
I) & & (B + 2\pi\alpha' F)_{\psi\theta} 
= 2 \kappa\alpha'(\phi\sin^2\psi\sin\theta + f')
\\\label{choiceII}
II) & & (B + 2\pi\alpha' F)_{\psi\phi} 
= 2 \kappa\alpha'(\sin^2\psi\cos\theta + f'') 
\end{eqnarray}
where $f'$ and $f''$ are undetermined parameters, which correspond to the 
$U(1)$ gauge field on D-brane worldvolume.
For the type I choice, the B-form potential in (\ref{choiceI})
has multivalue under the transformation $\phi\longrightarrow \phi + 2\pi$ 
which means it is an unphysical choice. So we turn to type II choice.
Inserting (\ref{choiceII}) into the action (\ref{action}), we find besides 
D0- and D1-branes described by the boundary conditions (\ref{d0})-(\ref{d1'}) 
additional D2-branes with the boundary conditions
\begin{eqnarray}\label{conic}
& &  \theta\Big{|}_{\partial\Sigma} = \theta_0 \nonumber
\\
 & & \(\partial_{\sigma}\psi - 
2(\sin^2\psi\cos\theta + f')\partial_{\tau}\phi\)\Big{|}_{\partial\Sigma} 
= 0 \nonumber
\\
 & & \(\sin^2\psi\sin^2\theta\partial_{\sigma}\phi + 
2(\sin^2\psi\cos\theta + f')\partial_{\tau}\psi\) \Big{|}_{\partial\Sigma} = 0
\end{eqnarray}
where (\ref{conic}) describes conic-like D2-branes. 
As we know, one of criteria to choose the B-form potential is 
that the dynamics disallows the end-points of strings
to hit the singularity\cite{1}. In type II case, there
is the conical singularity on type II D2-brane which
indicates that the choice (\ref{choiceII}) is
physically unacceptable. Thus in the spherical
coordinates the only physical 
choice for B-form potential is (\ref{bf}).
\enddel

\let\l=\lambda \let\r=\rho 
Since the group manifold is $O(4)$ symmetric, which manifests itself  
in the global symmetry of the action $g\to \l g\r$ under left- and 
right-multiplication with 
constant group elements, it is clear that there should also be (stable)
spherical D2-branes that are not centered around the unit element. Our 
coordinates are, of course, not very convenient for the discussion of these
objects, but it is obvious that our results carry over to that situation and
that they are related to the (inner) automorphisms of the current algebra
that were discussed in \cite{A,6}. Indeed, since 
$(rgr^{-1})hg^{-1}= r\l(hr^{-1})\l^{-1}= \r(hr^{-1})\r^{-1}r$
with $\l=g\r^{-1}$ and $\r=rgr^{-1}$
the twisted conjugay class defined by $h$ and the inner antomorphism 
corresponding to $r$ is just the sphere through $hr^{-1}$ centered around $r$.
In the exact CFT treatment it turns out that, at small levels \cite B,
the brane positions are somewhat smeared out.
It would be interesting to find out what the fate of the apparently 
unstable D1 branes is after quantum corrections are taken into account.

Eq.(\ref{sd2}) shows that the D2-brane sphere should be a fuzzy sphere. 
Indeed, there have been 
some discussions of noncommutative geometry on the spherical D2-branes with 
B-fields \cite{ARS, LM}. 
Especially in \cite{ARS} the low-energy effective action on the fuzzy $S^2$ 
was proposed and it would be %, then it is 
interesting to see whether there exists a similar Seiberg-Witten map 
\cite{SW} on the fuzzy sphere, and if so, how the 
nonlinear $\Lambda$-symmetry in noncommutative geometry is realized as in 
\cite{KZ}. As we know, among all examples 
in AdS/CFT correspondence, the boundary theory of $AdS_2 \times S^2$ is 
most poorly understood, see \cite{JGZ} for references. 
In \cite{HL} it was argued that besides the fuzzy $S^2$ there is also a 
fuzzy $AdS_2$. It would be interesting to see whether there is a way to 
study the fuzzy $AdS_2$ in the context of WZW models.

\bigskip\noindent{\large\bf Acknowledgement}
 
We would like to thank A.Y. Alekseev, N. Ishibashi, C. Schweigert,
V. Schomerus and S. Stanciu for helpful discussions. This work is 
supported in part by the {\it Austrian Research Funds} FWF under 
grants Nr. P13125-TPH and Nr. M535-TPH.

\end{document}